\documentclass[iop]{emulateapj}
\usepackage{graphicx}

\shorttitle{Magnetic Ribbon Model}
\shortauthors{Auddy et al.}

\begin{document}


\title{A Magnetic Ribbon Model for Star-Forming Filaments}

\author{Sayantan Auddy\altaffilmark{1}, Shantanu Basu\altaffilmark{1}, and Takahiro Kudoh\altaffilmark{2}}

\altaffiltext{1}{Department of Physics and Astronomy, The University of Western Ontario, London, ON N6A 3K7, Canada.}
\altaffiltext{2}{Faculty of Education, Nagasaki University, 1-14 Bunkyo-machi, Nagasaki 852-8521, Japan.}
\email{sauddy3@uwo.ca, basu@uwo.ca}
\email{kudoh@nagasaki-u.ac.jp}





\begin{abstract}
We develop a magnetic ribbon model for molecular cloud filaments. These result from turbulent compression in a molecular 
cloud in which the background magnetic field sets a preferred direction. We argue that this is a natural model for filaments and is based on the interplay between turbulence, strong magnetic fields, and  gravitationally-driven ambipolar diffusion, rather than pure gravity and thermal pressure.  An analytic model for the formation of magnetic ribbons that is based on numerical simulations is used to derive a lateral width of a magnetic ribbon. This differs from the thickness along the magnetic field direction, which is essentially the Jeans scale. We use our model to calculate a synthetic observed relation between apparent width in projection versus observed column density. The relationship is relatively flat, similar to observations, and unlike the simple expectation
based on a Jeans length argument.  

\end{abstract}

\newpage

\keywords{ISM: clouds --- magnetic fields --- magnetohydrodynamics (MHD) --- stars: formation --- turbulence}

\section{Introduction}
The {\it Herschel Space Observatory} has revealed a wide-ranging network of
elongated (filamentary) structures in molecular clouds
\citep[e.g.,][]{and10,men10}. Even though filamentary structures
in molecular clouds were already well established \citep[e.g.,][]{sch79},
the Herschel continuum maps of dust emission at
70-500 $ \rm{\mu} $m have achieved unprecedented sensitivity and revealed a
deeper network of filaments, in both star-forming and non-star-forming
molecular clouds. This implies that the filamentary network is an imprint
of initial conditions, likely turbulence, rather than the result of pure
gravitational instability. Furthermore, the prestellar cores and protostars,
when present, are preferentially found along massive filaments.

Much interpretation of the filaments has been based on the assumption that
they are isothermal cylinders. This simplifies their analysis as their
observed shape is then independent of most viewing angles and one can rely
on established theoretical results about the equilibrium or collapse of
infinite cylinders. \citet{and10} interpreted the observations in terms of
the critical line mass of an isothermal cylinder $m_{\rm l,crit} = 2\,c_s^2/G$,
where $c_s$ is the isothermal sound speed. For a mass per unit
length $m > m_{\rm l,crit}$, a cylinder undergoes
indefinite collapse as long as the gas is isothermal, and for
$m < m_{\rm l,crit}$ it can settle
into an equilibrium structure, although still unstable to clumping along its 
length into Jeans length sized fragments \citep{lar85}.
\citet{and10} argue that star formation is initiated when $m > m_{\rm l,crit}$.

A challenge to the view of filaments as cylinders is the
magnetic field alignment inferred from polarized emission.
\citet{pal13} find that large scale magnetic fields are
aligned perpendicular to the long axis of the massive star-forming
filaments \citep[see also][]{pla16}. This makes a circular symmetry of a cylinder about the long axis
unlikely unless the magnetic field strength is dynamically insignificant.
A more natural configuration is a magnetic ribbon, a triaxial object that
is flattened along the direction of the large-scale magnetic field
with its shortest dimension in that direction. In the lateral direction
to the magnetic field, elongated structures can form due to turbulence
and gravity. Indeed, simulations of turbulence accelerated star formation
in a strongly magnetic medium \citep{li04,nak05,kud08,bas09,kud11}
show the formation of ribbon-like
structure in a layer that is flattened along the magnetic field direction.
Magnetic ribbons have recently been investigated theoretically by
\citet{tom14} and \citet{han15}.  They study magnetohydrostatic equilibria 
of ribbons that arise from a parent filament of radius $R_0$, which 
is a free parameter in the problem. They find that a critical line-mass-to-flux
ratio exists for collapse, in analogy to the critical mass-to-flux ratio
for axisymmetric three-dimensional objects \citep{mou76}.

A further challenge to filaments modeled as isothermal cylinders comes from
the dust emission measurement of the FWHM of the mean column density profile relative
to the axis of a filament \citep{arz11}. For example, figure 7 of \citet{arz11}
shows that the FWHM values for 90 filamentary structures in low mass star forming 
regions cluster around a mean of
$\sim 0.1$ pc with some scatter over two orders of magnitude range of mean
column density\footnote{Molecular line emission studies of the Taurus region show
	wider mean thicknesses $\sim 0.4$ pc for 
	filaments in velocity-integrated emission and $\sim 0.2$ pc for filaments in 
        individual velocity channels \citep{pan14}}.
However, \citet{ost64} showed that the central half-mass
radius of an equilibrium isothermal cylinder is $a \propto c_s/\sqrt{G\rho_c}$,
essentially the Jeans length, where $\rho_c$ is the central density.
The projected column density of such a circularly symmetric configuration
has a central flat region of size $a$ and column density $\Sigma_c = 2 \rho_c\,a$
\citep[see][]{dap09}, so that we can also write $a \propto c_s^2/(G\Sigma_c)$.
Therefore, the approximate observed relation $a \simeq {\rm constant}$ is
unlike the expected $a \propto \Sigma_c^{-1}$.
However, the observed set of values of the FWHM radii also intersect the
line of Jeans length at the median log column density, which implies that the
Jeans length may not be wholly unrelated to them.

In this paper, we explore the consequences of a magnetic ribbon model for molecular cloud filaments for the measured relation between apparent width and the observed column density. We argue that this is a more natural model for filaments and is based on the interplay between turbulence, strong magnetic fields, and  gravitationally-driven ambipolar diffusion, rather than pure gravity and thermal pressure. We extend the analytic model of \citet{kud14} for the formation of magnetic ribbons that is based on numerical simulations. We derive a lateral width of a magnetic ribbon and use it to calculate a synthetic observed relation between apparent width in projection versus observed column density. 


\section{Semi-Analytic Model}

\subsection{Background}

Dynamically important magnetic fields, corresponding to mass-to-flux ratios that range from subcritical to mildly
supercritical, will lead to flattening along the magnetic field direction, and subsequent evolution will be primarily
perpendicular to the magnetic field \citep{fie93,nak08}. Even highly turbulent three-dimensional simulations 
\citep{kim13} show that the turbulence is eventually dominated by motions perpendicular to the ambient magnetic field.
Observations of some filaments \citep[e.g.,][]{pal13} that show a large-scale magnetic field along the short dimension
of the filament also support the idea of flattening along the field. In this paper, we adopt the scenario of 
turbulent compression acting primarily perpendicular to the magnetic field direction in an initially subcritical
molecular cloud. This leads to the paradigm of turbulence accelerated star formation, in which star formation
occurs with globally low efficiency and in turbulent compressed regions. These regions oscillate about an approximate
force-balanced state until ambipolar diffusion creates supercritical pockets that collapse to form stars.
We explore the consequences of this scenario by extending a semi-analytic model of \citet{kud14} that is based on numerical 
simulations.

\subsection{Ribbon Width} 

\begin{figure}[h]
	\label{ribbon}
\centerline{\includegraphics[height=8cm]{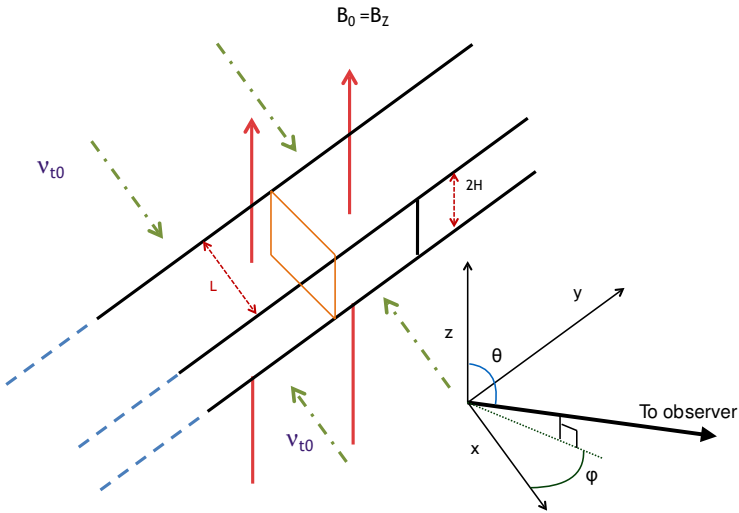}}
\caption{The formation of a magnetic ribbon as the molecular cloud contracts under the influence of the ``ram pressure" and the perpendicular magnetic field. The thick black arrow points to an observer located at a random orientation in the sky.  }
\end{figure}

We consider local pressure balance of a compressed region in a subcritical cloud and neglect thermal pressure in 
comparison to magnetic pressure and the ram pressure of the flow. We assume that the cloud is stratified in the 
\textit{z}-direction, with compression happening primarily in the $\textit{x-y}$ plane. Here we simplify the analysis 
of the compression by limiting it to one direction, the \textit{x}-axis (Fig. \ref{ribbon}), as done by \citet{kud14}. 
The initial magnetic field strength is $B_0$ and the field strength increases upon compression until the magnetic pressure 
within the compressed ribbon balances the external ram pressure and magnetic pressure. Hence the compression ends (and 
oscillations may ensue) when
\begin{equation}\label{balance eq}
H\frac{B^2}{8 \pi} = H_0 \left(\rho_0v_{t0}^2 +\frac{B_0^2}{8 \pi}\right),
\end{equation}
where $v_{t0}$ is the nonlinear flow speed. Assuming that the gas has adequate time to settle into hydrostatic equilibrium along the \textit{z}-direction, the half thickness of the cloud is 
\begin{equation}\label{scaleheight}
H =  \frac{c_s}{\sqrt{2 \pi G \rho}}
\end{equation}
\citep{spi42}. 
Now if the ambipolar diffusion time is longer than the compression time \citep{kud14}, flux freezing is valid during compression, i.e.,
\begin{equation}\label{fluxfreezing}
\frac{B}{\Sigma} = \frac{B_0}{\Sigma_0}.
\end{equation}
For the surface density $\Sigma = 2 \rho H$, equation (\ref{fluxfreezing}) can be rewritten as 
\begin{equation}\label{fluxfreezing1}
\frac{B}{\rho^{\frac{1}{2}}} =  \frac{B_0}{\rho_{0}^{\frac{1}{2}}}.
\end{equation}
Using equation (\ref{scaleheight}) and equation (\ref{fluxfreezing1}) in equation (\ref{balance eq}) and with some simplifications  we get
\begin{equation}\label{densityvelocity}
\left(\frac{\rho}{\rho_0}\right)^{1/2}= 2 \left(\frac{v_{t0}}{v_{A0}}\right)^{2}+1,
\end{equation} 
where $v_{A0}^{2}= B_0^2/(4\pi\rho_0)$ is the square of the initial Alfv\'en speed of the cloud. The consequence of such 
compression results in the formation of magnetic ribbons of width $L$ and thickness $2H$, as they are flattened along 
the direction of magnetic field (see Fig. \ref{ribbon}). 
For conservation of mass per unit length in the ribbon during the compression of the cloud
\begin{equation}\label{mass conservation}
\rho_0L_0H_0 =\rho L H,
\end{equation}
where $L_0$ is the initial width (along the \textit{x}-axis) and $2H_0$ is the initial thickness of the cloud in the 
vertical direction (\textit{z}-axis). 
Using equation (\ref{scaleheight}), we can simplify the above equation to 
\begin{equation}\label{densitylength}
\left(\frac{\rho}{\rho_0}\right)^{1/2} = \frac{L_0}{L}.
\end{equation}
By using equation (\ref{densitylength}) in equation (\ref{densityvelocity}), we can express the final width of 
the filament as 
\begin{equation}
L= L_0\left[2 \left(\frac{v_{t0}}{v_{A0}}\right)^{2}+1\right]^{-1}.
\end{equation}
Analysis of Zeeman measurements of the magnetic field in molecular clouds presented by \citet{cru99} shows that 
the turbulent line width is comparable to the Alfv\'en speed \citep{bas00}. 
If we make the plausible estimate that the flow speed is comparable to the Alfv\'en speed, i.e., $v_{t0} \simeq v_{A0}$, 
the filament width becomes
\begin{equation}\label{Intrinsic width}
L \simeq L_0/3.
\end{equation}
The result illustrates the fact that the final width of a filament is {\it independent} of the density of the medium. Instead it is a fraction of the initial length scale $L_0$ of the compressed region.  

\subsection{Initial Compression Scale}\label{LO}

In the above theory, the final ribbon width $L$ is independent of its density, but does depend on the initial compression
scale $L_0$ associated with turbulence. The origin and physics of $L$ is then quite different than that of the Jeans
length. What is $L_0$ then?  In the turbulent scenario that we adopt in 
this study, it would be associated with the dominant mode in the turbulent flow field in a molecular cloud.  At this point, no first principles theory
exists to calculate $L_0$ as the preferred mode of an instability that leads to molecular cloud turbulence. Hence, we take guidance from observations
to make an empirical estimate for $L_0$.

\subsubsection{Estimate from column density map}

 The column density maps used by \citet{arz11} to determine the mean filament width can also be used to estimate a mean spacing between filaments, which 
we identify with $L_0$ in our model. Figure 3b of \citet{arz11} identifies 27 filaments in a dust emission map of the cloud IC 5146. We use this same map
to make an approximate measurement of $L_0$. For each of the 27 filaments, 
we identify a center along the length (spine) of the filament. Then for
each filament we measure the distance to the nearest center point of another
filament. We obtain a set of 20 unique distance measurements (eliminating
double counting in cases where two filaments are mutually each other's
nearest neighbor). 
We ignore effects of an inclination angle $i$ in this analysis, which 
could mean that the measured distances are less than the actual distances 
by a factor $\sin i$.
Our measured filament spacings (which we equate with $L_0$) have a 
minimum value 0.5 pc, a maximum
value 2.2 pc, a median value 0.9 pc, and a mean value 1.0 pc. 

\subsubsection{Estimate from star formation timescale}

Another way to constrain $L_0$ is through the star formation timescale in molecular clouds. Since this number is widely accepted to be in the range 1-3 Myr \citep {pal00,pal02,har01}, and star 
formation is often coordinated along a filament over this timescale, we can place an upper 
limit (again empirically) on the compression timescale $t_0$ of a few Myr.
Therefore the initial length scale that can trigger a compression can be written as
\begin{equation}
L_0 \simeq  v_{t0}\, t_0 \simeq v_{A0}\,t_0,
\end{equation} 
where $t_0$ is as above and again using the Alfv\'enic nature of turbulence.
This further simplifies to 
\begin{equation}
	L_0 \simeq \frac{B_0}{\sqrt{4\pi \rho_0}}\, t_0 = \frac{\sqrt{2}}{\mu_0}\,c_s\,t_0,
\end{equation}
where we have used pressure balance along the magnetic field, $\pi G \Sigma_0^2/2 =
\rho_0\,c_s^2$ and the normalized mass-to-flux ratio $\mu_0 = \Sigma_0(2 \pi \sqrt{G})/B_0$.  
If we consider the initial cloud to be mildly subcritical, i.e., $\mu_0 \approx 0.5$, 
and a sound speed $c_s = 0.2 \,\rm  km\, s^{-1} \simeq 0.2 \,\rm  pc\, Myr^{-1}$,
then $t_0 \simeq 1-3$ Myr leads to $L_0 \simeq 1$ pc.

\section{Results}
The simple arguments of the previous section show that the length scale 
	at which the ribbon formation is initiated is 
of the order of a parsec. Our semi-analytic model then implies that
the final width of the ribbon given by equation (\ref{Intrinsic width}) is $\sim 0.3$ pc.
However, the shortest dimension is flattened along the direction of the magnetic field and has a thickness $2H$ that
{\it does} depend on the column density. Therefore, the observed shape will depend on the viewing angle. Below we calculate the observed width for a particular viewing angle and then calculate a synthetic plot of observed ribbon width versus observed column density for a collection of random viewing angles. Our objective is to gain insight into the form of the observed correlation, and how it compares with the standard Jeans length scaling and with the observational results presented by \citet{arz11}. The value of $L_0$ can be considered a free parameter and physically may vary from one cloud to another and have a distribution of values within a single cloud. While we do not advocate a specific individual value for $L_0$, we use the empirical estimate that it should be $\sim 1$ pc to determine the shape and approximate quantitative values of an observed correlation. 

\subsection{Observed Width}
Let the normal to the filament, along the \textit{z}-axis,  be inclined at an angle $\theta$ to the observer as shown in Fig. \ref{orientation}. For a ribbon-like filament of intrinsic width $L$ and half thickness $H$ the projected width $L_{\rm obs}$ as seen by the observer is
\begin{equation}\label{L_observed}
L_{\rm obs} = L \cos\theta + 2H \sin \theta.
\end{equation}
If the ribbon is viewed face on, i.e., $\theta =0^{\circ} $, the observed width is just the intrinsic width $L$. When viewed side on i.e., $\theta =90^{\circ} $ the observed width is the thickness $2H$ of the ribbon along the \textit{z} axis. For any other intermediate angles one sees the projection in the \textit{y-z} plane i.e., equation (\ref{L_observed}), as shown in Fig. \ref{orientation}.\\
From our analysis we have already shown in equation (\ref{Intrinsic width}) that the intrinsic width $L$ is a fraction of the initial compression length scale $L_0$. 

The thickness $2H$ of the magnetic ribbon  is evaluated using the hydrostatic equilibrium, equation (\ref{scaleheight}), along the direction of the magnetic field (i.e., perpendicular to the filament width). For column density $\Sigma = 2 \rho H$, the half-thickness of the clouds is estimated to be
\begin{equation}\label{Hsigma}
H = \frac{c_s^2}{\pi G \Sigma}.
\end{equation}
For a ribbon of any particular column density $\Sigma$ we can estimate the corresponding half-thickness $H$, which is essentially the Jeans scale, using equation (\ref{Hsigma}). 
For example, $H = 0.16$ pc for $c_s = 0.2 \,\rm{km\, s^{-1}}$ and $N \equiv \Sigma/m = 10^{21} \rm{cm^{-2}}$ in which $m = 2.3\,m_{\rm H}$.

\subsection{Observed Column Density}

The observed column density $\Sigma_{\rm obs}$ will be different from the intrinsic column density $\Sigma$ depending on the angle at which the ribbon is being viewed. If the observer is situated at angle other than $\theta = \phi= 0^{\circ}$ (as shown in  Fig. \ref{ribbon}), the length along the line of sight changes thus affecting the observed column density. In the following section we will analyze the variation of $\Sigma_{\rm obs}$ with the viewing angles $\theta$ and $\phi$.  We neglect the variation of $\rho$ within the ribbon. 

\subsubsection{Case 1 ($0^{\circ} \leq\theta \leq\theta_{\rm crit})$}
For the beam incident on the face of the ribbon at an angle $0^{\circ}\leq\theta \leq \theta_{\rm crit}$ and $\phi =0 ^{\circ}$, (refer to Fig. \ref{orientation}a,) the observed column density is

\begin{equation}
\Sigma_{\rm obs} = 2\rho  H \sec \theta.
\end{equation}
Since the intrinsic column density  $\Sigma = 2 \rho H$, we get


\begin{equation}\label{sigma1}
\Sigma_{\rm obs} = \Sigma  \sec \theta.
\end{equation}
Thus only for $\theta = 0 ^{\circ}$, i.e., when the ribbon is viewed face on, $\Sigma_{\rm obs} = \Sigma$. For $0^{\circ}<\theta\leq\theta_{\rm crit}$, $\Sigma_{\rm obs} > \Sigma$.

\subsubsection{Case 2 ($ \theta=\theta_{\rm crit}  $)}

\begin{figure}[h]
\includegraphics[height=13cm]{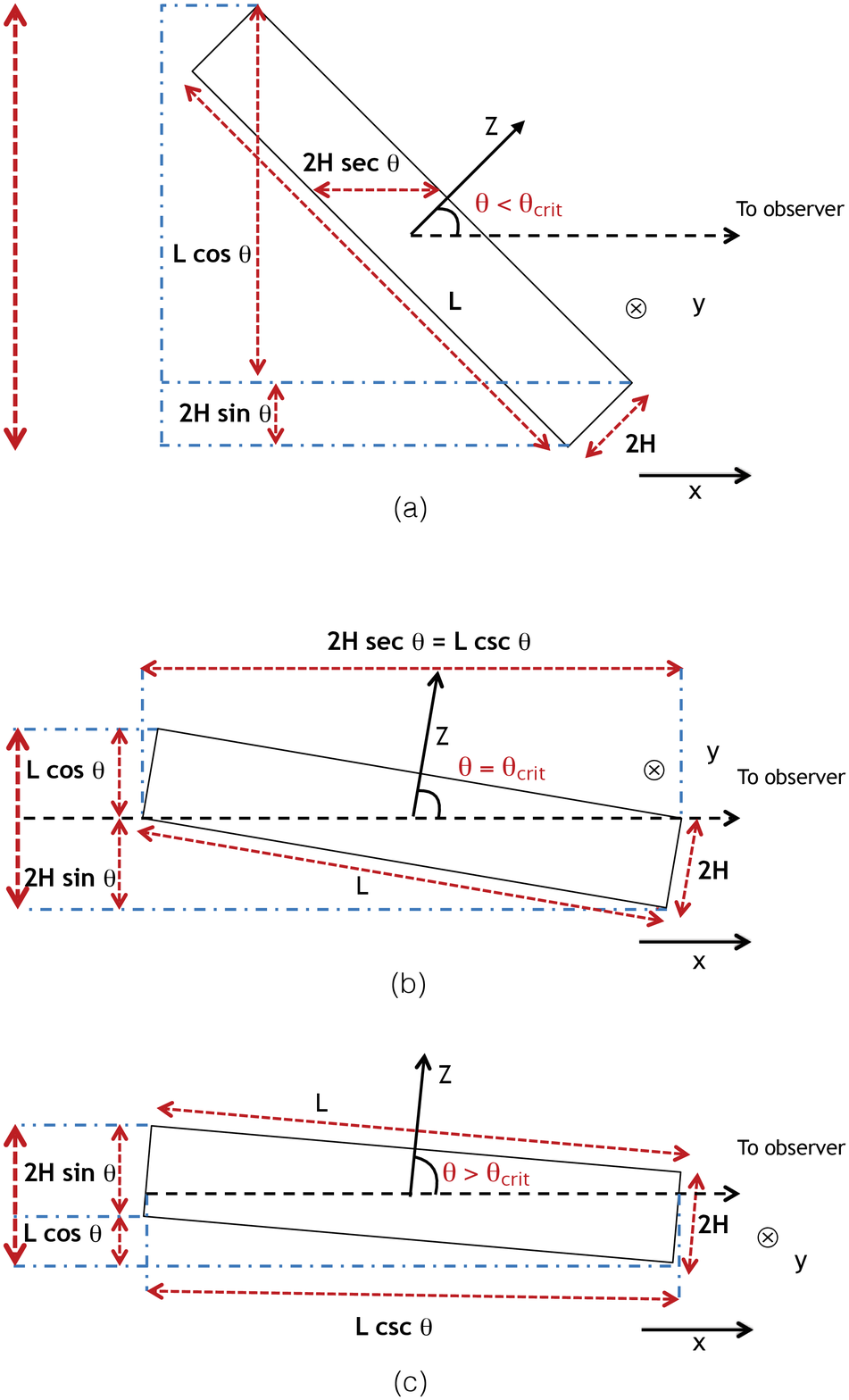}
\caption{Different orientations of the magnetic ribbon with respect to the observer. Top panel: case 1 ($0^{\circ} \leq\theta \leq\theta_{\rm {crit}}$). Middle panel: case 2 ($ \theta=\theta_{\rm crit}  $) when the ribbon is observed at a critical angle as shown. Bottom panel: case 3 ($ \theta_{\rm crit}\leq\theta \leq 90^{\circ}  $). }\label{orientation}
\end{figure}

For the beam entering at a critical angle $\theta_{\rm crit}$ and $\phi =0 ^{\circ}$ (Fig. \ref{orientation}b), the observed column density is

\begin{equation}
\Sigma_{\rm obs} = 2 \rho H \sec\theta_{\rm crit} = \rho  L \csc \, \theta_{\rm crit}.
\end{equation}
Rearranging the above equation, we  find that
\begin{equation}
\theta_{\rm crit} = \tan^{-1}\frac{L}{2H}.
\end{equation}

The critical angle $\theta_{\rm crit}$ separates the two sets of angles that have separate expressions for $\Sigma_{\rm{obs}}$.

\subsubsection{Case 3 ($ \theta_{\rm crit}\leq\theta \leq 90^{\circ}  $)}
For the beam entering through the shorter dimension of the ribbon (see Fig. \ref{orientation}c)  at an angle $\theta_{\rm crit} \leq \theta \leq 90^\circ $ and $\phi =0 ^{\circ}$, the observed column density  is   
\begin{equation}
\Sigma_{\rm obs} = \rho L \csc \, \theta.
\end{equation}
Using equation (\ref{scaleheight}), and $\Sigma = 2 \rho H$, we get

\begin{equation}\label{sigma2}
\Sigma_{\rm obs} = \frac{\Sigma L}{2H\sin \theta} = \frac{\pi G \Sigma^2 L}{2c_{s}^{2}\sin \theta }.
\end{equation}
For $\theta = 90^{\circ} $, i.e., when the ribbon is viewed side on, $\Sigma_{\rm obs} = \Sigma\,\frac{L}{2H}$. For $\theta_{\rm crit}<\theta \leq 90^\circ $, $\Sigma_{\rm obs} > \Sigma$ if $L > 2H $, i.e., when the ribbon width is greater than the thickness of the ribbon, the observed column density is greater than the actual column density.

\subsubsection{Case 4 ($\phi \neq 0^{\circ}$)}

Furthermore, different angular orientation of the magnetic ribbons in the \textit{{x-y}} plane will also alter the observed column density. If the long axis (\textit{y}) of the ribbon is not perpendicular the line of sight (see Fig. \ref{ribbon}), i.e., $\phi \neq 0^{\circ}$ the observed column density will further increase. For any random orientation in the \textit{{x-y}} plane, the modified column density is
\begin{equation}
\Sigma_{\rm obs}= \Sigma  \sec \theta \sec \phi ,\,\, 0^{\circ} \leq\theta \leq\theta_{\rm crit},
\end{equation}
\begin{equation}
\Sigma_{\rm obs}=   \frac{\pi G \Sigma^2 L}{2c_{s}^{2}\sin \theta }\sec \phi ,\,\,\theta_{\rm crit}\leq\theta \leq 90^{\circ}. 
\end{equation}

However, different orientations in the \textit{{x-y}} plane  do not affect the observed ribbon width.  The resultant projection of the ribbon width on the \textit{{y-z}} plane is independent of the azimuthal angle $\phi$. 

\subsection{Observed Correlation}
\label{obscorr}
\begin{figure}[h]
\includegraphics[height=5.5cm]{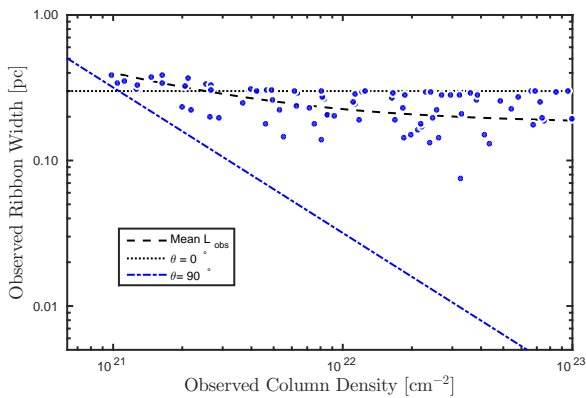}
\caption{Apparent ribbon width $L_{\rm obs}$ versus observed column density $N_{\rm obs}$. Each blue dot corresponds to a magnetic ribbon with intrinsic column density $N$ and observing angle $\theta$, chosen as described in Sec. \ref{obscorr}. The black dashed line is the mean ribbon width for the entire range of values of $N_{\rm obs}$. The black dotted line is the width when the ribbon is viewed at $\theta = 0 ^{\circ}$. The blue dot-dashed line is the width for the side on view i.e., $\theta = 90 ^{\circ}$.   } \label{scatter}
\end{figure}

Since the observations of \citet{arz11} reveal a relatively flat relation between observed width
and column density, it is instructive to use our model to make a synthetic map of these
quantities. For simplicity we consider $\phi=0$ in this analysis. We take a sample of 100 ribbons
with number column density distributed uniformly in the range $10^{21}\,\rm{cm^{-2}}\leq \textit{N} \leq10^{23}\,\rm{cm^{-2}}$. Furthermore we 
take viewing angles randomly chosen in the range $0^{\circ} \leq\theta\leq 90^{\circ}$. Each pair
of values $(N,\theta)$ yields a pair of values $(N_{\rm obs},L_{\rm obs})$ represented as blue
dots in Fig. \ref{scatter}. We obtain $L_{\rm obs}$ using equation (\ref{L_observed}) and $\Sigma_{\rm obs}$ using equation (\ref{sigma1}) or equation (\ref{sigma2}) for $\theta \leq \theta_{\rm crit}$ or $\theta >  \theta_{\rm crit}$, respectively. The black dashed line in Fig. \ref{scatter} is the locus
of points obtained by taking 100 randomly chosen values of $\theta$ for each value of $N$ and 
calculating the mean values
of $L_{\rm obs}$ and $N_{\rm obs}$. This line is similar to the result of 
taking equation (\ref{L_observed}) and inserting the mean values of $\cos \theta$ and $\sin \theta$, 
which are both equal to $2/\pi$, and replacing $\Sigma$ with the mean value of  $\Sigma_{\rm obs}$. However,  
the mean value of $\Sigma_{\rm obs}$ across all angles is not exactly equal to $\Sigma$.
Both the set of individual synthetic data points shown in blue dots as well as the average relation 
in the black dashed line show a relatively flat relation over two orders of magnitude variation in
$N_{\rm obs}$.
Fig. \ref{scatter} also shows the analytic relation for two limiting cases. 
The black dotted line corresponds to the face on view ($\theta = 0 ^{\circ}$) where 
$L_{\rm obs} = L = 0.3$ pc and is independent of the column density. 
The blue dot-dashed line corresponds to $\theta = 90 ^{\circ}$ where $L_{\rm obs} = 2c_s^2/ (\pi G\Sigma)$, essentially the Jeans length. 

\section{Conclusion}
We have presented a minimum hypothesis model for the width of
a filament in a molecular cloud in which magnetic fields and 
magnetohydrodynamic turbulence are initially dominant. 
A turbulent compression leads to 
a magnetic ribbon whose thickness is set by the standoff between
ram pressure and magnetic pressure region. 
Gravitationally-driven ambipolar diffusion then leads to runaway
collapse of the densest regions in the ribbon, where the mass-to-flux
ratio has become supercritical.
 This process has been demonstrated
in published simulations of trans-Alfv\'enic turbulence in a cloud 
with an initial subcritical mass-to-flux ratio 
\citep[e.g.,][]{nak05,kud11}.   
We have extended the semi-analytic model of \citet{kud14} to estimate
their lateral (perpendicular to magnetic field and ribbon long axis)
width. This quantity is independent of the density of the ribbon.
 This lateral width can also be used to estimate the parent filament radius
$R_0$ in the theoretical magnetic ribbon model of \citet{tom14}.
In our model, the thickness parallel to the magnetic field is essentially
the Jeans scale and does depend on density. Hence, we calculate a  
distribution of apparent widths seen in projection assuming a random
set of viewing angles. The resulting distribution of apparent widths
versus apparent column density is relatively flat (unlike expectations
based on the Jeans length) over the range $10^{21}$ cm$^{-2}$ -- 
$10^{23}$ cm$^{-2}$, in rough agreement with the observations of 
\citet{arz11}.  
Other models have been introduced to explain the apparent near-uniform
width of observed filaments. 
\citet{fis12} introduce an external pressure to an isothermal cylinder and
find that the FWHM versus column density is a peaked function and 
approximately flat in the regime where $m_{\rm l}$ is well below 
$m_{\rm l,crit}$ and the external pressure is 
comparable to the central pressure. 
However, filaments with $m_{\rm l} > m_{\rm l,crit}$
would be in a time-dependent state of dynamical collapse.
 \citet{hen13} develop a model of a cylindrical
self-gravitating filament that is accreting at a prescribed rate. 
A near-uniform radius is derived based on assumption of a 
steady-state balance between energy input from accretion and
dissipation of energy by ion-neutral friction at the filament radius scale. 
 \citet{hei13} also develops a model of accretion at the 
free-fall rate onto a filament with $m_{\rm l} < m_{\rm l,crit}$ and 
uses various prescribed forms of internal structure to find that 
the FWHM has a peaked dependence on column density.
A series of simulation papers
\citep{smi14,kir15,fed16} use hydrodynamic or MHD simulations (with
supercritical mass-to-flux ratio)
and analyze filament widths at particular snapshots in time. 
Although their filament widths cluster at $\sim 0.1$ pc with some scatter, there is a mild
to strong density dependence of the widths, and the filaments
are single time snapshots in a situation of continuing collapse. \citet{fed16}
suggests that $\sim 0.1$ pc is special since the linewidth-size relation 
of \citet{lar81} would lead to subsonic turbulence below that scale, but it 
is not clear if his simulations satisfy this scaling internally.
 We believe that the magnetic ribbon model provides an alternative 
simplified interpretation that accounts for turbulence and strong magnetic
fields. We have developed a method to estimate the width of a magnetic
ribbon based on the characteristic scale and amplitude of MHD turbulence. 
Such ribbons can have a line mass that exceeds the hydrodynamic limit
$2 c_s^2/G$ and still be in a dynamically oscillating quasi-equilibrium 
state. However, gravity still leads to star formation in the dense interior
through rapid ambipolar diffusion. 
\acknowledgments
We thank the anonymous referee for comments that improved the discussion
in this paper. This work was supported by a Discovery Grant from NSERC.

\newpage


\end{document}